\documentclass[psfig,a4paper,12pt,leqno,newlfont]{article}
\usepackage[latin1]{inputenc}
\usepackage{a4}
\usepackage{epsfig}
\date{}
\title{1/r - POTENTIAL WITHOUT CHARGE}
\author{K. Buchner}

\begin{document}
\maketitle

\vspace{1cm}

\begin{abstract}
In order to get geodesically complete Rei{\ss}ner-Nordstr{\o}m space-times, it is necessary to identify pairs of singular points. This can be done in such a way that "wormholes" \/ are created which generate electric field lines without any charge. Finally, it is shown that it is possible to glue this space-time not in the singularities $r=0$, but at some $r>0$. The surface energy generated by this gluing is exotic, but tends to zero in the limit $r \rightarrow 0$. 
\end{abstract}

\section{Introduction}

More than 40 years ago, Archibald Wheeler has realized \cite{Whe1}, \cite{Whe2} that non-trivial electric fields can be generated, whose field lines do not end in electric charges. Instead some domain of the space is cut out, in which the charge would be expected. To this "hole", a channel ("wormhole") is attached, through which the field lines pass to some other place such that the divergence of the field lines vanishes. So no charges are needed - they are replaced by special topological structures. \\

For the realization of this idea, one starts from the Rei{\ss}ner-Nordstr{\o}m solution \cite{Nord}, \cite{Rei} to Einstein's equations. It contains an electric field proportional to $1/r^2$. Visser has cut out the regions $r<\epsilon$ from two Rei{\ss}ner-Nordstr{\o}m space-times, and has glued together the remaining parts along the subspaces $r=\epsilon$ \cite{Vis1}, \cite{Vis2}. He wanted to obtain a "transversable" wormhole. Therefore he has chosen $\epsilon$ larger than the outer horizon $r_+$. But this causes some problems: \\

The geodesics passing through the subspace $r=\epsilon$ are not $C^2$. (Trivially, they can be made continuously differentiable everywhere.) This means that freely falling particles get a $\delta$-like kick at  $r=\epsilon$. Second, the Ricci tensor contains a singularity proportional to $\delta(r-\epsilon)$. Therefore the energy-momentum tensor $T$ has surface terms. They violate the "average null energy condition" 
\begin{eqnarray} 
g_{ij} \, \dot{\gamma}^i(\tau) \, \dot{\gamma}^j(\tau) = 0 \qquad \Rightarrow \qquad \int T_{ij} \, \dot{\gamma}^i(\tau) \, \dot{\gamma}^j(\tau) \, d\tau \ge 0 \label{Gleich00} \\
\tau \mbox{: generalized affine parameter} \qquad \quad \; \nonumber
\end{eqnarray}
\cite{HawEll}, \cite{Vis2} for some curves $\tau \, \mapsto \, \gamma(\tau)$, i.e. the surface matter is "exotic". This is true for all spherically symmetric transversable wormholes, whatever the specific assumptions are. A good survey of the literature on this subject is given in \cite{Vis2}. \\

In the present work, we want to develop a model for the topological generation of charge via wormholes. This is done for the specific example of electric charge, but, of course, the idea applies to all kinds of charge. For this purpose, it is not necessary to have transversable wormholes: We do not need information from the "other part of the world". In addition, we can not even expect to get transversable wormholes, because in the Rei{\ss}ner-Nordstr{\o}m solution with $M^2>Q^2$, the mass and charge are hidden by two horizons. So it would be very surprising, if one could avoid all horizons by introducing wormholes and leave the rest of physics unchanged.\\

The most natural way to reach our goal is to identify singularities $r=0$ in the maximal analytic extension of the Rei{\ss}ner-Nordstr{\o}m solution. Then in the case $M^2>Q^2$, no changes of space-time outside the horizons are needed. By this identification, the topology becomes non-trivial - a fact which is to be expected: Einstein´s equations determine only the local geometry, but in most cases give little information about the global structure of space-time. It has to be determined by other requirements. \\

In order that this identification makes sense, one needs a precise definition of singularities. This is still an open problem. But here, we are not interested in the most general situation. Instead, one can start from the maximal analytical extention and use some theory defining the necessary geometric quantities in the singularity. The details of such a theory are irrelevant in this connection, because all functions are uniquely determined by their behaviour near the singularity. In the present work, the Rei{\ss}ner-Nordstr{\o}m solution together with its singularities is considered as a d-space \cite{GeBu1}, \cite{GeBu2}. They are almost identical to the differential spaces of \cite{HeS} and \cite{Mos}. For the discussion of differential equations on these spaces, the results of \cite {BuB} are used. But as remarked above, most other generalizations of differential manifolds will lead to the same results. This is true in particular for the differential spaces described in \cite{HeS} and \cite{Mos}.\\

In addition, gluing of two Rei{\ss}ner-Nordstr{\o}m solutions at some  $r=\epsilon > 0$ will be discussed. Here, exotic surface matter appears. But the energy of the electromagnetic field rises as $1/r$, whereas the contribution of the surface matter tends to zero for $\epsilon \to 0$. Therefore, by an appropriate choice of $\epsilon$, the ratio of the surface energy to the electric energy can be made arbitrarily small. \\

Gluing at $r=0$ has the advantage that all geodesics are everywhere $C^3$. On the other hand, gluing at some $\epsilon > 0$ avoids the singularities in the metric and the electric field, but entails surface matter.

\section{The global structure of the Rei{\ss}ner-Nordstr{\o}m solution}

The Rei{\ss}ner-Nordstr{\o}m solution \cite{Nord}, \cite{Rei} to Einstein's equations
\begin{displaymath}
R_{ik} \, - \, \frac{1}{2} \, R \, g_{ik} = \frac{8 \, \pi \, G}{c^2} \, T_{ik} \qquad G \mbox{: Newton's gravitational constant}
\end{displaymath}
describes a static spherically symmetric object with electric and magnetic fields. It is given by the metric
\begin{equation} \label{Gleich2.1}
ds^2 = \frac{\Delta}{r^2} dt^2 - \frac{r^2}{\Delta} dr^2 - r^2 \, d \Omega^2 
\end{equation}
with the abbreviations
\begin{displaymath}
\Delta = r^2 - 2 M r + Q^2
\end{displaymath}
and
\begin{displaymath}
d \Omega ^2 = d \vartheta ^2 + sin^2 \vartheta \, d\varphi^2 \; .
\end{displaymath}
Here $M$ and $Q$ are constants determined by the mass $m$, the electric charge $q$ and the magnetic charge $q_m$, respectively, of the object:
\begin{displaymath}
M = \frac{2 \, G \, m}{c^2} \, ; \qquad Q^2 = \frac{G}{4 \, \pi \, \epsilon_0 \, c^4} \, (q^2 + q_m^2)
\end{displaymath}
So the Rei{\ss}ner-Nordstr{\o}m solution may contain also a magnetic monopole. But in the following, we will set $q_m=0$, unless the contrary is stated explicitely. \\

We shall restrict ourselves to the case $M^2 > Q^2$. For elementary particles, this condition is violated by many orders of magnitude. On the other hand, all "elementary" objects, i.e. quarks and leptons, have spin. Therefore the Rei{\ss}ner-Nordstr{\o}m solution does not apply to them, no matter, wether $M^2 > Q^2$ or $M^2 < Q^2$. \\

The construction of the maximal analytic extention starts from the three regions
\begin{eqnarray*}
A &:& 0 < r < r_- \\
B &:& r_- < r < r_+ \\
C &:& r_+ < r < \infty \ .
\end{eqnarray*}
Here $r_+$ and $r_-$ are the zeros of $\Delta$:
\begin{eqnarray}
r_+ & := & M \, + \, \sqrt{M^2 \, - \, Q^2} \\
r_- & := & M \, - \, \sqrt{M^2 \, - \, Q^2} \; . \label{Glei201}
\end{eqnarray}
It is convenient to introduce new coordinates $u$, $v$ by
\begin{eqnarray}
\label{Gleich2.5}  A \mbox{ and } C: \qquad &u& := t \, - \, r_*; \qquad v := t \, + \, r_* \\
\label{Gleich2.6} B: \qquad  &u& := t \, + \, r_*; \qquad v := r_* \, - \,t, 
\end{eqnarray}
with the abbreviation
\begin{equation} \label{Gleich26a}
r_* := \int \frac{r^2}{\Delta} dr = r + \frac{r_+^2}{r_+ - r_-} \, ln \mid r - r_+ \mid \, - \, \frac {r_-^2}{r_+ - r_-} \, ln \mid r - r_- \mid
\end{equation}
Next, one constructs the new regions $A^\prime, \; B^\prime$ and $C^\prime$ from $A, \; B$ and $C$, respectively, by replacing $u \mapsto -u; \quad v \mapsto -v$ in (\ref{Gleich2.5}) and (\ref{Gleich2.6}). These six regions are composed to a periodic "ladder" according to fig. 1 (see, e.g. \cite{Chand}, \cite{HawEll}). The transformations
\begin{eqnarray}
A_n:  U &:=& arc \, tg(-e^{-\alpha \, u}) \; + \; (n + 1) \, \pi \nonumber \\
      V &:=& arc \, tg(e^{\alpha \, v}) \; + \; n \, \pi \nonumber \\
B_n:  U &:=& arc \, tg(e^{\alpha \, u}) \; + \; n \, \pi \nonumber \\
      V &:=& arc \, tg(e^{\alpha \, v}) \; + \; n \, \pi \nonumber \\
C_n:  U &:=& arc \, tg(-e^{-\alpha \, u}) \; + \; n \, \pi \nonumber \\
      V &:=& arc \, tg(e^{\alpha \, v}) \; + \; n \, \pi \nonumber \\
A^\prime_n: U &:=& arc \, tg(-e^{-\alpha \, u}) \; + \; (n-\frac{1}{2}) \, \pi \label{Gleich26bb}\\ 
            V &:=& arc \, tg(e^{\alpha \, v}) \; + \; (n-\frac{1}{2}) \, \pi \nonumber \\ 
B^\prime_n: U &:=& arc \, tg(e^{\alpha \, u}) \; + \; (n-\frac{1}{2}) \, \pi \nonumber \\
            V &:=& arc \, tg(e^{\alpha \, v}) \; + \; (n-\frac{1}{2}) \, \pi \nonumber \\
C^\prime_n: U &:=& arc \, tg(-e^{-\alpha \, u}) \; + \; (n+\frac{1}{2}) \, \pi \nonumber \\
            V &:=& arc \, tg(e^{\alpha \, v}) \; + \; (n-\frac{1}{2}) \, \pi \nonumber \\
\alpha := (r_+ \! \! \! \! & - & \! \! \! \!  \, r_-) \, / \, (2\, r_+^2) \nonumber 
\end{eqnarray}
define global parameters $(U, \, V, \, \vartheta, \, \varphi)$. Here $U+V$ may be interpreted as "time", and $U-V$ as "radius". With these coordinates, the line element (\ref{Gleich2.1}) becomes:
\begin{equation} \label{Gleich2.10}
ds^2 = - \frac{4}{\alpha ^2 \, r^2} \, \frac{\mid r \, - \, r_+ \mid \, \mid r \, - \, r_- \mid }{sin \, ( 2 \, U) \; sin \, (2 \, V)} \; dU \, dV \, - \, r^2 \, d\Omega^2 \; .
\end{equation} \\

\begin{figure}[htb]
\begin{center}
\epsfig{file=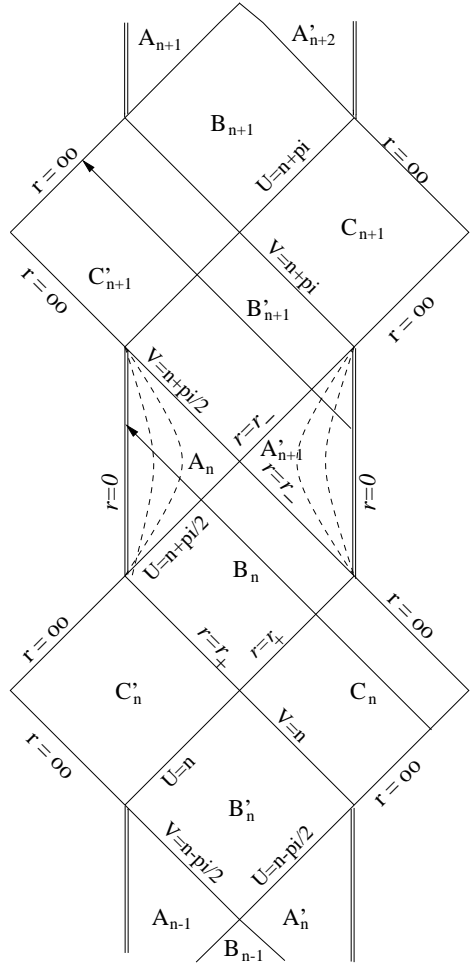, scale=.6}
\caption{Maximal analytic extention of the Rei{\ss}ner-Nordstr{\o}m space-time. Dashed lines: r=const. The solid line with arrows shows a photon starting in $C_n$ and passing to $C_{n+1}^\prime$ through a singularity.} 
\label{Bild1}
\end{center}
\end{figure}

It can be seen from fig.\ref{Bild1} that this maximal analytic extension is not geodesically complete, because all radial spacelike and lightlike geodesics end in the singularities. This suggests to identify the singularities in $A_n$ with those in $A^\prime_m$ for suitable pairs $m, \, n$. But, of course, the metric (\ref{Gleich2.10}) and its derivatives are not defined in $r=0$. So it is not possible to apply here the usual rules \cite{Lan}, \cite{Sen} for gluing spaces in General Relativity, i.e. to require that the metric (\ref{Gleich2.10}) is at least $C^2$ in the tangential directions and $C^1$ in the normal direction. Instead, one consideres the geodesics passing  through the singularities. By definition, they need to be $C^2$-functions of their parameters. And in order to define "geodesic" coordinates, it is necessary \cite{BeB} that their limits for $r \rightarrow 0$ are at least $C^3$ in $A_n \cup A_{m}^\prime$. But in general it is not possible to work with affine parameters, as the metric is singular at $r=0$. \\ 

The first step is to generalize differential manifolds in such a way that the singularities can be included. For this, the d-spaces of \cite{GeBu1}, \cite{GeBu2} will be used, which have been very useful in the discussion of Schwarzschild´s metric, the radiation filled Friedmann universe and some more exotic space-times \cite{ABG}, \cite{Bu1}. The basic idea is to replace the coordinate functions in the definition of differential manifolds by a more general set $C$ of functions. In particular, no compatibility conditions between different charts are needed. The only requirement is that one can add, substract and multiply the functions in $C$. This leads to the following definition: \\

Let $M$ be a topological space. The pair $(M, C)$ is called d-space, if $C$ is a sheaf of continuous real-valued functions on $M$ which form an algebra w.r.t. pointwise operation. \\

For these d-spaces, it is simple to construct the tangent vectors as "directional derivatives": \\

Let $(M, C)$ be a d-space and $C_x$ the stalk at $x \in M$. A map
\begin{displaymath}
V: C_x \rightarrow I \! \! R
\end{displaymath}
is called tangent vector to $(M, C)$ in $x$, if for all $n \in I \! \! N$, all $f_1, \dots , f_n \in C_x$, and all germs $\alpha$ of $C^1(I \! \! R^n, I \! \! R)$ at $y := (f_1(x), \dots , f_n(x)) \in I \! \! R^n$, the equation
\begin{displaymath}
V \left( \alpha \circ \left( f_1(x), \dots , f_n(x) \right) \right) = \sum \limits_{i=1}^n \; (\partial_i \, \alpha) \cdot V(f_i)
\end{displaymath}
holds, provided $\alpha \circ (f_1, \dots , f_n) \in {\cal C}_x$. Here $\partial_i \, \alpha$ denotes the partial derivative of $\alpha$ w.r.t. the $i$-th argument. \\

For the construction of geodesics, it is necessary to have some theory of differential equations. In general, such a theory does not exist. But in the case under discussion, the regular points are dense in the d-space. Here the results of \cite{BuB} guarantee existence and uniqueness of the solutions. \\

In such a theory it is possible to glue the singular points in $A_n$ to those in $A_m^\prime$. It is natural to require $m=n+1$ and to identify points with equal "local times" $t$ or with equal "global times" $U+V$. We choose the latter possibility: If the points $(r=0, t) \in A_n$ are identified with the points $(r=0, t) \in A^\prime_{n+1}$, there exist causal geodesics, which come arbitrarily close to themselves. To see this, regard the lines in fig.\ref{Bild1} with $U=n+\pi /2$ or $V=n+\pi /2$. A lightlike geodesic just below such a line would be continued to a geodesic just above that line. \\

Gluing points with $r=0$ and equal values of $U+V$ entails a time reflexion, i.e. an identification of the points $(r=0, t)$ in $A_n$ with the points $(r=0, -t)$ in $A_{n+1}$. \\

\section{The geodesics in $r=0$}

The geodesics are most easily computed in the local coordinates  $(r, \, \vartheta, \, \varphi, \, t)$. The method is similar to that used for Schwarzschild's solution: Because of spherical symmetry, it is possible to put $\vartheta = \pi/2 = const$. In this way, the equations are considerably simplified, and the first integrals are easily obtained. The results for spacelike geodesics are \cite{Chand}:
\begin{eqnarray}
\label{Gleich2.11} \frac{d \varphi}{d \tau} & = & \frac{L }{r^2} \\
\label{Gleich2.12} \frac{dt}{d \tau} & = & E \, \frac{r^2}{\Delta} \\
\label{Gleich2.13} \left( \frac{dr}{d \tau} \right)^2 & = & E^2 \, + \, \frac{\Delta}{r^2} \, \left(1 \, - \, \frac{L^2}{r^2} \right) \quad .
\end{eqnarray}
$E$ and $L$ are constants of integration. It can be seen from (\ref{Gleich2.13}) that spacelike geodesics in $A_n$ and in $A^\prime_m$ reach the value $r=0$, if and only if $L=0$ (radial geodesics). In the limit $r \rightarrow 0$, equations (\ref{Gleich2.11}) - (\ref{Gleich2.13}) yield for $L=0$ 
\begin{equation} \label{Gleich2.14}
\frac{d}{d\tau} \left( r(\tau), \, \vartheta(\tau), \, \varphi(\tau), \, t(\tau) \right) \, \rightarrow \, \left( \pm  \frac{Q}{r}, \, 0, \, 0, \, E \, \frac{r^2}{Q^2}\right)
\end{equation}
or
\begin{equation} \label{Gleich2.15}
\quad \; \left( r(\tau), \, \vartheta(\tau), \, \varphi(\tau), \, t(\tau) \right) \, \rightarrow \, \left( \pm \sqrt{2 \, \mid Q \, \tau \mid }, \, \frac{\pi}{2}, \, \varphi_0, \, sign(\tau) \, \frac{E}{|Q|} \, \tau^2 + t_0 \right) \; .
\end{equation}

These equations show that the radial spacelike geodesics can be continued through the singularities such that they are (at least) $C^3$ in the sense of d-spaces. This includes the fact that they are three times continuously differentiable on $A_n \cup A_{n+1}^\prime - \{(r=0, t)\, | \, t \in I \! \! R \}$. But this is only true, if suitable curve parameters $\sigma$ (not the arc length $\tau$, but e.g. $\sigma := \tau/r$) and appropriate constants of integration are chosen (positive sign in (\ref{Gleich2.15}) and positive (resp. negative) sign in (\ref{Gleich2.14}) for $Q\,\tau > 0$ (resp. $ < 0$); $E$ and $t_0$ change their sign in $\tau=0$.) Notice that in $\sigma=0$ both, $dr/d\sigma$ and $t(\sigma)$ have to change sign. Otherwise the geodesics are not $C^3$. A general discussion of the gluing conditions can be found in the review paper \cite{BeB}. \\

Of course, the arc length $\tau$ is not a good parameter near $r=0$, because the metric is singular and $\tau$ is not well defined. If the tangent points in $r$-direction, i.e. if $E=0$,
equation (\ref{Gleich2.13}) yields:
\begin{equation} \label{Gleich2.16}
\frac{dr}{d\tau} = \pm \sqrt{\frac{\Delta}{r^2}}
\end{equation}
with the solution (remember that $\Delta \ge 0$ holds in $A$ and $A^\prime$):
\begin{displaymath}
\sqrt{\Delta} \, + \, M \, ln \left( M - r - \sqrt{\Delta} \right) = \pm \tau \, + \tau_0 \; .
\end{displaymath}
Here $\tau_0$ is a constant of integration. This implicit equation for $r(\tau)$ shows that the r-lines are geodesics which can be continued through the singularities. \\

In addition to the spacelike geodesics, also the radial lightlike geodesics reach the singulatity. Their equations can be derived as in the spacelike case \cite{Chand}. With an affine parameter $\tau$ they read:
\begin{eqnarray}
\frac{d r}{d \tau} & = & \pm E \nonumber \\
\frac{d t}{d \tau} & = & \frac{E \, r^2}{\Delta} \label{Gleich2001} \\
\frac{d \vartheta}{d \tau} & = & \frac{d \varphi}{d \tau} = 0  \nonumber \; .
\end{eqnarray}
This yields
\begin{displaymath}
\frac{d r}{d t} = \pm \frac{\Delta}{r^2}
\end{displaymath}
with the solution
\begin{displaymath}
t = \pm r_* + konst 
\end{displaymath}
(cf.(\ref{Gleich26a})). It is convenient to choose the parameter $\tau$ such that $r(0)=0$ holds. If in addition the constant $E$ changes sign in $r=0$, then these geodesics are of class $C^\infty$ in $r=0$ (in the local coordinates $U$ and $V$ modulo $\pi/2$; cf. (\ref{Gleich26bb}) and \cite{Bei}). \\

\section{The charge in the Rei{\ss}ner - Nordstr{\o}m solution}

The electromagnetic field tensor of the Rei{\ss}ner-Nordstr{\o}m solution is \cite{Vis2}:
\begin{equation} \label{Gleich3.1}
F = \frac{1}{4 \, \pi \, \epsilon_0} \left( \frac{q}{r^2} \frac{\partial}{\partial \, t} \wedge \frac{\partial}{\partial \, r}  + \frac{q_m}{r^4 \, sin \, \vartheta} \frac{\partial}{\partial \, \vartheta}  \wedge \frac{\partial}{\partial \, \varphi}  \right)
\end{equation}
From this, the components $j^i$ of the four current can be computed:
\begin{equation} \label{Gleich3.2}
j^i = F^{ik}_{;k} \; .
\end{equation}
Here the semicolon denotes the covariant derivative. For $r>0$, it follows immediately from (\ref{Gleich4.1}) below, that the components $j^1$ and $j^2$ vanish. The other components are:
\begin{displaymath}
j^3 = \frac{-1}{4 \, \pi \, \epsilon_0} \left[ \frac{d}{d\vartheta} \left( \frac{q_m}{r^4 \, sin \, \vartheta}  \right) + \frac{q_m}{r^4 \, sin \, \vartheta} \, cot \, \vartheta   \right] = 0
\end{displaymath}
and
\begin{displaymath}
j^4 = \frac{q}{4 \, \pi \, \epsilon_0} \left( \frac{2}{r^3} \, - \, \frac{2}{r^3} \right) = 0 \;  .
\end{displaymath}
At $r=0$, equation (\ref{Gleich3.2}) can not be used, as $F$ diverges. But it is possible to compute the exterior derivative by Gau{\ss}'s theorem. This means to discuss the electric field lines near $r=0$. In our coordinates, these are the $r$-lines. It has been shown in the last section that they pass through the singularities. Therefore the divergence of the electric field vanishes: After gluing the space-time, the $\epsilon$-neighbourhood $V$ of the points $r=0$ consists of two balls with center at $r=0$: One in $A_n$, and one in $A_{n+1}^\prime$. Consider 
\begin{eqnarray} \label{Gleich3.3}
\int\limits_V J &=& \int\limits_Vd\,F^* = \int\limits_Vd(\epsilon_{ijkl} \, F^{kl} \, dx^i \wedge dx^j) \nonumber \\
&=& \int\limits_{\partial \, V} \epsilon_{ijkl} \, F^{kl} \, dx^i \wedge dx^j 
\end{eqnarray}
with $q_m=0$. Then only the components $(k, \, l)=(1, \, 4)$ contribute. As the field lines crossing the sphere in $A_n$ pass it from outside to inside, while they pass the sphere in $A_{n+1}^\prime$ in the opposite direction, the total surface integral over $\partial \, V$ vanishes. This is even true, if $q_m\ne0$: The term $F^{23}$ in (\ref{Gleich3.3}) is present only in integrals in $x^1$- and $x^4$-direction. But in the $x^1$-integration, the contribution from $A_n$ cancels that from $A_{n+1}^\prime$. So also this integral vanishes, and we finally get:
\begin{displaymath}
\int\limits_V J = 0 \; .
\end{displaymath}

\section{Energy and momentum}

In this section, we glue the space-time not in the singularity $r=0$, but at finite values of $r$. For this, we choose some $\epsilon$ satisfying $0<\epsilon<r_-$ (see (\ref{Glei201})) and delete all points with $r<\epsilon$. Then we identify the points $(r=\epsilon, t)$ in $A_n$ with the points $(r=\epsilon, -t)$ in $A_{n+1}^\prime$ in a similar way as we did before. The stability of such surfaces $r=\epsilon$ between two parts of Rei{\ss}ner - Nordstr{\o}m space-times has been discussed by Visser \cite{Vis1}, \cite{Vis2}. \\

The $r$-lines are geodesics perpendicular to the boundary $r=\epsilon$. If one follows such a line, the $r$-values first decrease, until the boundary $r=\epsilon$ is reached. Then they increase again. But, of course, at the boundary $dr/d\tau \ne 0$ (cf. (\ref{Gleich2.16})). Therefore the derivatives of the metric in the direction of $\tau$ are discontinuous and the Einstein tensor gets additional terms proportional to $\delta(r-\epsilon)$ \cite{Lan}, \cite{Sen}. The existence of such terms can also be seen from a simple consideration: If they would be absent, the radial lightlike geodesics would go on to the singularity $r=0$. But at $r=\epsilon$, they have to change their direction towards increasing $r$-values. Therefore some interaction is needed which forces them to do this. \\

In the coordinates $(r, \, \vartheta, \, \varphi, \, t)$, the only non-vanishing Christoffel symbols are:
\begin{eqnarray} 
\Gamma^1_{11} & = & \frac{1}{2} \; \frac{-\frac{2 \, M}{r^2} + \frac{2 \, Q^2}{r^3}}{1 \, - \, \frac{2 \, M}{r} \, + \, \frac{Q^2}{r^2}} \nonumber \\
\Gamma^1_{22} & = & - \left( 1 \, - \, \frac{2 \, M}{r} \, + \, \frac{Q^2}{r^2} \right) \, \cdot \, r \nonumber \\
\Gamma^1_{33} & = & - \left( 1 \, - \, \frac{2 \, M}{r} \, + \, \frac{Q^2}{r^2} \right) \, \cdot \, r \, sin^2 \, \vartheta \nonumber \\
\Gamma^1_{44} & = & -\frac{1}{2} \; \left( 1 \, - \, \frac{2 \, M}{r} \, + \, \frac{Q^2}{r^2} \right) \, \left( -\frac{2 \, M}{r^2} + \frac{2 \, Q^2}{r^3} \right) \label{Gleich4.1}\\
\Gamma^2_{12} & = & \Gamma^2_{21} \, = \, \frac{1}{r} \nonumber \\
\Gamma^2_{33} & = & -\, sin \; \vartheta \, cos \, \vartheta \nonumber \\
\Gamma^3_{13} & = & \Gamma^3_{31} \, = \, \frac{1}{r} \nonumber \\
\Gamma^3_{23} & = & \Gamma^3_{32} \, = \, cot \, \vartheta \nonumber \\
\Gamma^4_{14} & = & \Gamma^4_{41} \, = \, -\frac{1}{2} \; \frac{-\frac{2 \, M}{r^2} + \frac{2 \, Q^2}{r^3}}{1 \, - \, \frac{2 \, M}{r} \, + \, \frac{Q^2}{r^2}} \nonumber 
\end{eqnarray}
The normal to the boundary, pointing away from the region $A$ resp. $A^\prime$ has the components
\begin{equation} \label{Gleich4001}
(n_i) = \left( \frac{-1}{\sqrt{1 \, - \, 2 \, M/\epsilon \, + \, Q^2/\epsilon^2}}, \, 0, \, 0, \, 0 \right); \qquad \quad n^i \, n_i = -1 \; .
\end{equation}
The second fundamental form $K$ of the boundary is defined by
\begin{displaymath}
K_{ij} = - \frac{1}{2} \left( (\nabla_i \, n)_j \, + \, (\nabla_j \, n)_i \right) \; ; \qquad i,j=2,3,4 \; .
\end{displaymath}
Together with (\ref{Gleich4.1}), this yields in the limit $r \rightarrow \epsilon; \; r < \epsilon$:
\begin{eqnarray}
K_{22} &=& \epsilon \, \sqrt{1 \, - \, 2 \, M/\epsilon \, + \, Q^2/\epsilon^2} \nonumber \\
K_{33} &=& \epsilon \, sin^2 \vartheta \; \sqrt{1 \, - \, 2 \, M/\epsilon \, + \, Q^2/\epsilon^2} \label{Gleich4.2} \\
K_{44} &=& - (M/\epsilon^2 - Q^2/\epsilon^3) \; \sqrt{1 \, - \, 2 \, M/\epsilon \, + \, Q^2/\epsilon^2} \nonumber
\end{eqnarray}
All other components of $K$ vanish. Therefore the energy-momentum tensor $T$ gets the additional terms from the boundary \cite{Vis2}
\begin{displaymath}
Z_{ij} = \frac{c^2}{4 \, \pi \, G} \, \delta(r-\epsilon) \, \left(-K_{ij} \, + \, K^r_r \, g_{ij} \right)\, ; \qquad i,j,r=2,3,4 \; .
\end{displaymath}
For $K^r_r$, (\ref{Gleich4.2}) yields
\begin{displaymath}
K^r_r = \frac{-2+3M/\epsilon-Q^2/\epsilon^2}{\epsilon \, \sqrt{1-2M/\epsilon+Q^2/\epsilon^2}} \; .
\end{displaymath}
In the limit $\epsilon \rightarrow 0$, the leading terms in $K^r_r$ are $-|Q|/\epsilon^2$. So one obtains
\begin{equation} \label{Gleich4.3}
Z_{44} \to \frac{-c^2 \, \delta(r-\epsilon)}{4 \, \pi \, G} \; \frac{2 \, |Q|^3}{\epsilon ^4}  \; .
\end{equation}
The additional energy due to the gluing is
\begin{displaymath}
E = \frac{1}{6} \int u^r \, Z_r^i \, \epsilon_{ijkl} \, dx^j \, dx^k \, dx^l \; ,
\end{displaymath}
where $\epsilon_{ijkl}$ is defined as
\begin{displaymath}
\epsilon_{ijkl} = \sqrt{| det \, g |}\left\{ \begin{array}{r} +1 \\ 0 \\ -1 \end{array} \right.
\end{displaymath}
and $u$ is the four-velocity of an observer at rest:
\begin{displaymath}
(u^r) = (0, \, 0, \, 0, \, \frac{1}{\sqrt{1 - 2M/\epsilon + Q^2/\epsilon^2}}) \; .
\end{displaymath}
(\ref{Gleich4.3}) shows that the additional energy $E$ is negative and tends to zero, if $\epsilon$ goes to zero. Therefore there is exotic matter (i.e. matter violating the average null energy condition) on the surface $r=\epsilon$, but its contribution to the total energy can be made arbitrarily small: Remember that the components $T^{(em)\, i}_{\qquad j}$ of the electromagnetic energy-momentum tensor are proportional to $1/\epsilon^4$ (see (\ref{Gleich3.1})). So we obtain the well-known result that the electromagnetic energy $E^\prime$ is proportional to $1/\epsilon$, i.e. the ratio between the electromagnetic energy and the surface energy becomes arbitrarily large. Therefore also the stability of the system is completely determined by the electromagnetic self-energy. \\

To sum up, we have the following results: Gluing the two parts of the Rei{\ss}ner-Nordstr{\o}m solution at some small $r = \epsilon > 0$ avoids the infinite self-energy of a point charge and the singularities in the metric. Furthermore, although there exists exotic surface matter, its contribution to the total energy can be made arbitrarily small, if $\epsilon$ tends to zero.  On the other hand, if the two parts of the Rei{\ss}ner-Nordstr{\o}m solution are glued together in the singularity $r = 0$, there is no surface energy.

\vspace{2cm}
\hspace{2cm} K. Buchner \\
\mbox{} Energiepolitischer Sprecher der \"odp \\
 Stra{\ss}bergerstr.16, D-80809 M\"unchen

\end{document}